\documentclass[11pt,twoside]{article}
\usepackage{graphicx,epsfig,natbib,epstopdf}
\usepackage{CS18}
\begin{document}
%
%
%
\title{Stars with and without planets: Where do they come from?}
%
%
\author{V.~Zh.~Adibekyan$^{1}$, J.~I.~Gonz\'{a}lez Hern\'{a}ndez$^{2,3}$, E.~Delgado~Mena$^{1}$, S.~G.~Sousa$^{1,2,4}$, N.~C.~Santos$^{1,4}$, 
G.~Israelian$^{2,3}$, P.~ Figueira$^{1}$, S. Bertran de Lis$^{2,3}$}
\affil{$^1$Centro de Astrof\'{\i}ísica da Universidade do Porto, Rua das Estrelas, 4150-762 Porto, Portuga}
\affil{$^2$Instituto de Astrof\'{\i}sica de Canarias, 38200 La Laguna, Tenerife, Spain}
\affil{$^3$Departamento de Astrof{\'\i}sica, Universidad de La Laguna, 38206 La Laguna, Tenerife, Spain}
\affil{$^4$Departamento de F\'{\i}ísica e Astronomia, Faculdade de Ci\^{e}ncias da Universidade do Porto, Portugal}

\begin{abstract}
%
%
A long and thorough investigation of chemical abundances of planet-hosting stars that lasted for more than a decade has finally beared fruit.
We explore a sample of 148 solar-like stars to search for  a possible correlation between the slopes of the abundance trends versus 
condensation temperature (known as the T$_{c}$ slope) both with stellar parameters and Galactic orbital parameters in order to understand the nature of 
the peculiar chemical signatures of these stars and the possible connection with planet formation. 
We find that the T$_{c}$ slope correlates at a significant level (at more than 4$\sigma$)  with the stellar age and the stellar surface gravity. 
We also find tentative evidence that the T$_{c}$ slope correlates with the mean galactocentric distance of 
the stars (R$_{mean}$), suggesting that stars that originated in the inner Galaxy have fewer refractory elements relative to the volatile ones.
We found that the chemical “peculiarities” (small refractory-to-volatile ratio) of planet-hosting stars is 
merely a reflection of their older age and their inner Galaxy origin.
We conclude that the stellar age and probably Galactic birth place are key to establish the abundances of some specific elements.
\end{abstract}
%
%
%
%
%
\section{Introduction}

Dozen of studies during the last decade explored the connection between stellar and
planetary properties. Naturally, this connection is found to be bidirectional: stellar properties play an
important role on planet formation and evolution (e.g. stellar metallicity-giant planet frequency - 
\citet{1997MNRAS.285..403G,2001A&A...373.1019S}), and the planet formation may have an impact on stellar 
properties (e.g. extra depletion of lithium in planet-hosting stars - \citet{2009Natur.462..189I, 2014A&A...562A..92D}).

After the first planets discovered, astronomers have been also trying to understand if the stars hosting planets are chemically peculiar 
(in terms of individual elements) and even to search for chemical signatures of planet formation on the hosting stars atmospheres.
For the first part the most significant result was recently obtained by \citet{2012A&A...543A..89A, 2012A&A...547A..36A}
who found that most of the metal-poor planet hosts are enhanced in $\alpha$-elements. For the second part 
(chemical imprints of planet formation) the results are still feeding a lively debate.

Several studies suggested that the chemical abundance trend with the condensation temperature, T$_{c}$, is a signature of 
terrestrial planet formation \citep[e.g.][]{2009ApJ...704L..66M, 2009A&A...508L..17R}. In particular, that the Sun shows "peculiar" chemical 
abundances because of the presence of the terrestrial planets in our solar system \citep{2009ApJ...704L..66M}.
Although these conclusions have been strongly debated in other studies \citep[e.g.][hereafter GH10,13]{2010ApJ...720.1592G, 2013A&A...552A...6G}, 
the main reason of the observed chemical “peculiarities” was not identified.

Here we explore the origin of the trend observed between [X/H] (or [X/Fe]) and T$_{c}$ using a sample of 148 solar-like stars from 
GH10,13. The more detailed analysis and complete results are presented in \citet{2014A&A...564L..15A}.

\section{Data}

Our initial sample is a combination of two samples of solar analogs (95 stars) and ``hot'' analogs (61 stars) taken from 
GH10,13. We have cross-matched this sample with the Geneva-Copenhagen Survey sample \citep[GCS-][]{2004A&A...418..989N}, for which \citet{2011A&A...530A.138C}
provides the Galactic orbital parameters, the space velocity components, and the ages of 148 of the stars considered in our study%
\footnote{Throughout the paper, BASTI expectation ages are used as suggested by \citet{2011A&A...530A.138C}.%
}. Fifty-seven of these stars are planet hosts, while for the remaining 91 no planetary companion has been detected up to now.

The stellar atmospheric parameters and the slopes of the $\Delta$[X/Fe]$_{SUN-star}$ versus  T$_{c}$ were derived using very high-quality 
HARPS spectra%
\footnote{Zero slope means solar chemical composition, and a positive slope corresponds to a smaller refractory-to-volatile ratio compared to the Sun.%
}. Twenty-five elements from C (Z = 6) to Eu (Z = 63) have been used for this analysis. 
These slopes are corrected for the Galactic chemical evolution trends as discussed in GH10,13.  

The stars in the sample have effective temperatures 5604 \emph{K} $\leq$ T$_{eff}$ $\leq$ 6374 \emph{K}, 
metallicites -0.29 $\leq$ {[}Fe/H{]} $\leq$ 0.38 dex, and surface gravities
4.14 $\leq$ $\log g$ $\leq$ 4.63 dex. Throughout the paper we defined solar analogs as stars with; T$_{eff}$ = 5777$\pm$200 K; 
logg = 4.44$\pm$0.20 dex; [Fe/H] = 0.0$\pm$0.2 dex. Fifteen out of 58 solar analogs in this sample are known to be orbited by planets.

\section{Correlations with T$_{c}$ slope}

We searched for possible correlations between the T$_{c}$ slope and, in turn, atmospheric parameters, and also Galactic orbital parameters
and age, in order to understand which is/are the main factor(s) possibly responsible for the abundance trends with T$_{c}$.

\subsection{T$_{c}$ slope against stellar parameters and age}

After a detailed analysis, we found that the T$_{c}$ trend strongly relates (at more than 4$\sigma$) with the surface gravity and stellar age (see Figure 1): old
stars are more “depleted” in refractory elements (smaller refractory-to-volatile ratios) than their younger
counterparts. At the same time we found no significant correlation of the T$_{c}$ slope with other stellar parameters.

Since for FGK dwarf stars in the main sequence one does not expect significant changes in their
atmospheric chemical abundances with age, we are led to believe that the observed correlation is likely to
reflect the chemical evolution in the Galaxy.
We note that this is the simplest assumption we can make based on our limited current knowledge of
stellar evolution, and we caution the reader that there might be other effects that could severely affect the composition of stars as a
function of age .

\subsection{T$_{c}$ slope and Galactic orbital parameters}

Moving one step further, we found a tentative evidence that the T$_{c}$ slopes correlate also
with the mean galactocentric distance of the stars (R$_{mean}$), which we use as a proxy of the birth radii (see Figure 2)%
\footnote{Several studies have shown that the mean of the apo- and pericentric distances, R$_{mean}$, are good indicators of the stellar birthplace
\citep[e.g.][]{1987JApA....8..123G, 1993A&A...275..101E}%
}. This trend is indicating that stars which have originated in the inner Galaxy have less refractory elements relative to the volatiles. 
This result qualitatively agrees with the recent observations of Galactic abundance gradients by \citet{2013A&A...558A..31L},
where the authors used young Galactic Cepheids for the gradient derivations.  

\subsection{Tc slope and planets}

Following our definition of solar analogs, we found that the average of the T$_{c}$ slope for planet hosting solar analogs is 
greater (0.012$\pm$0.31) than that of their non-host counterparts (-0.16$\pm$0.34). The Kolmogorov-Smirnov (K-S) statistics 
predict the $\approx$ 0.21  probability ($P_{KS}$) that these two 
subsamples came from the same underlying distribution for T$_{c}$ slope. At the same time, the same statistics predict a 
$P_{KS}$ $\approx$ 0.20 probability that they stem from the same underlying age distributions.
The latter can be seen in Figure 1: most of the planet-hosting stars tend to be relatively old ($>$ 5 Gyr).
Moreover,  planet host and non-host samples show a different distribution of  R$_{mean}$ -- $P_{KS}$ $\approx$ 0.007.
As can also be seen in Figure 2, 66\% planet hosts have R$_{mean}$ smaller than 7.5 kpc (where
slopes are usually high) and only 37\% of stars without detected planets have similarly low R$_{mean}$ values.
Clearly the two subsamples are not consistent with respect to the mean galactocentric distance and age.
Interestingly, \citet{2009ApJ...698L...1H} has already shown that (giant) planet host stars tend to have smaller R$_{mean}$ and probably 
originate in the inner disk, which follow the same direction as our findings.

These results suggest that the difference in T$_{c}$ slopes observed for solar analogs with and without planets is then probably due to the 
differences in their ``birth places'' and birth moment.

\section{Conclusion}

Our findings lead us to two interesting conclusions i) The solar analogues with planets in
the solar neighborhood mostly come from the inner Galaxy (because of still unknown reason) and ii) the age
and galactic birth place are the main factors responsible for the abundance ratio of refractory to volatile elements in
the stars.

\begin{figure}
\plotone{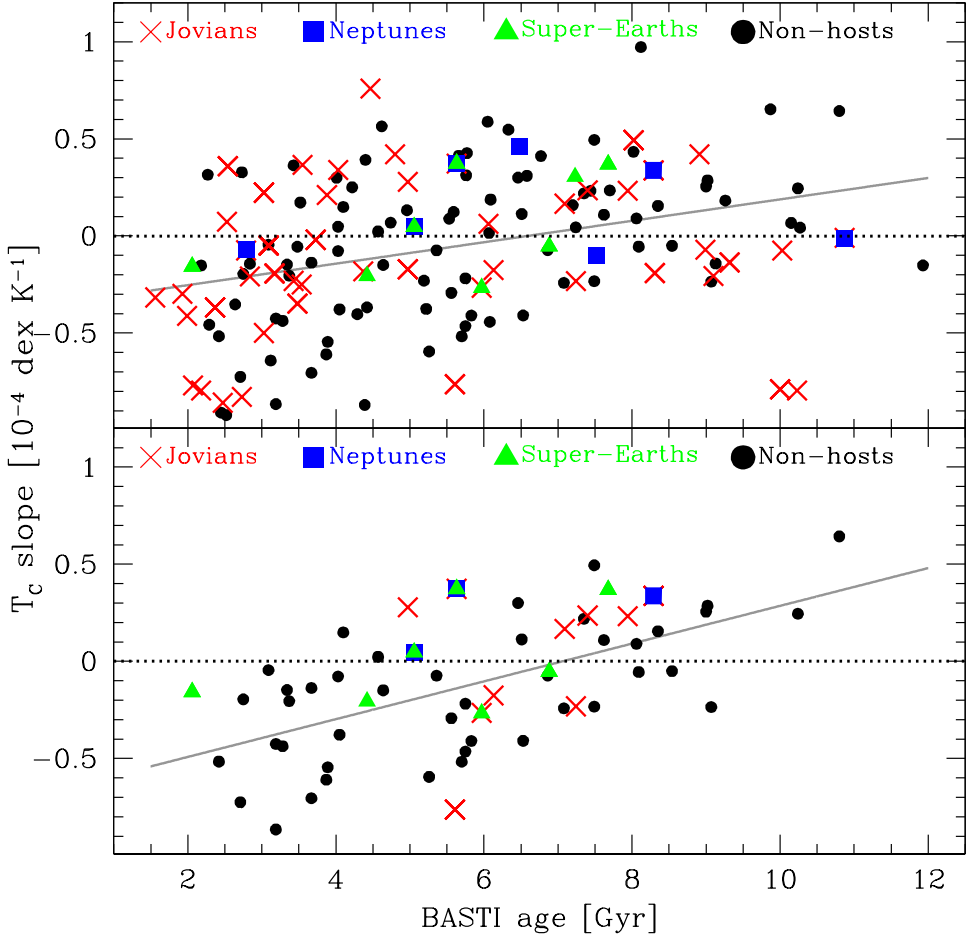}
\caption{T$_{c}$ slopes versus ages for the full sample (\textit{top}) and for the solar analogs (\textit{bottom}).
Gray solid lines provide linear fits to the data points.}
\end{figure}

\begin{figure}
\plotone{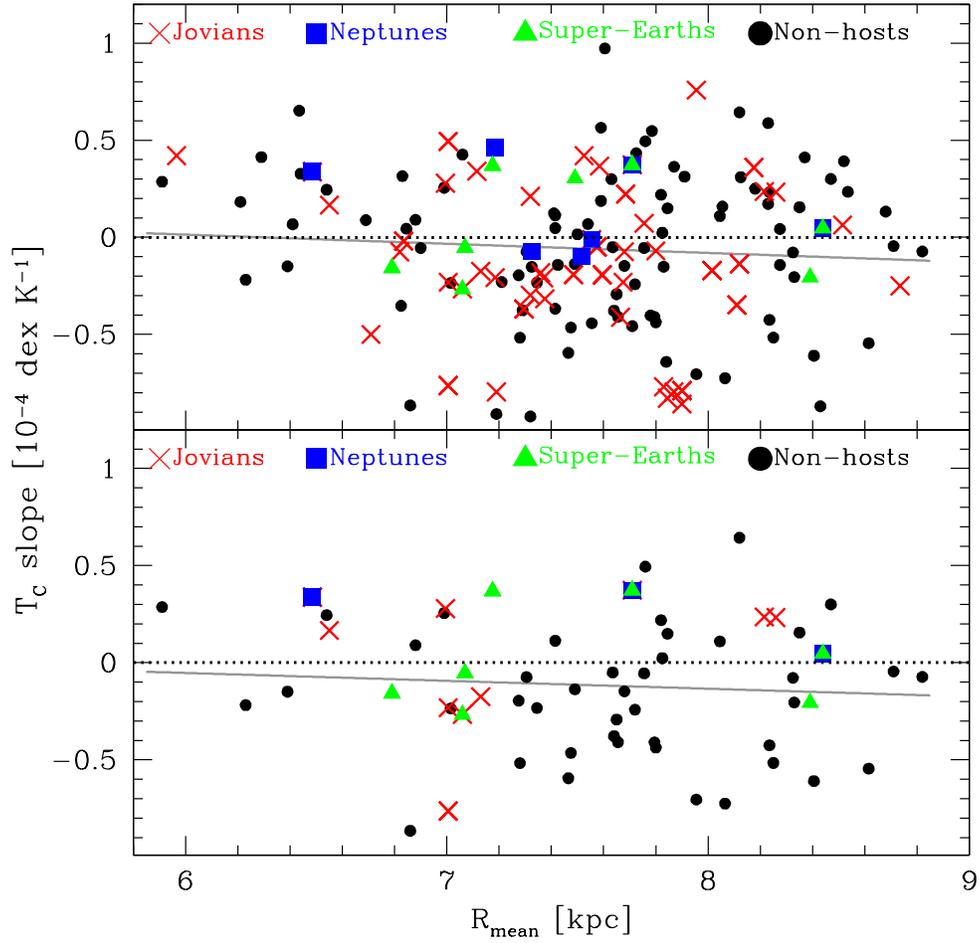}
\caption{T$_{c}$ slopes versus R$_{mean}$ for the full sample (top) and for the solar
analogs (bottom). Gray solid lines provide linear fits to the data points.}
\end{figure}

%
%

%
%

%
%

\acknowledgments{
This work was supported by the European Research Council/European Community under the FP7 through Starting Grant agreement 
number 239953. V.Zh.A., S.G.S., and E.D.M are supported by grants SFRH/BPD/70574/2010, 
SFRH/BPD/47611/2008, and SFRH/BPD/76606/2011 from the FCT (Portugal), respectively.
NCS also acknowledges support in the form of a Investigador FCT contract funded by FCT/MCTES (Portugal) and POPH/FSE (EC).
G.I., S.B.L, and J.I.G.H. acknowledge financial support from the Spanish Ministry project MINECO AYA2011-29060, 
and J.I.G.H. also received support from the Spanish Ministry of Economy
and Competitiveness (MINECO) under the 2011 Severo Ochoa Program MINECO SEV-2011-0187.
PF is supported by the FCT and POPH/FSE (EC) through an Investigador FCT contract with application reference IF/01037/2013 and
POPH/FSE (EC) by FEDER funding through the program "Programa Operacional de Factores de Competitividade - COMPETE.
}

\end{document}